\documentclass[runningheads]{llncs}

\usepackage{graphicx}
\usepackage[hidelinks]{hyperref}%Added to hide the boxes around links and citations 
\usepackage{booktabs}
\usepackage{footnote}
\usepackage{listings}
\usepackage{multirow}
\usepackage{footmisc}
\usepackage{subcaption}

\usepackage{booktabs}
\usepackage{tabularx,ragged2e}
\usepackage{array}
\usepackage{multirow}     	
\usepackage{longtable}		

\begin{document}
\title{Question Answering on Scholarly Knowledge Graphs}
%\title{Question Answering on Scholarly Tables with JarvisQA}
%
%\titlerunning{Question Answering on Scholarly Knowledge Graphs}
\author{Mohamad Yaser Jaradeh\inst{1}\orcidID{0000-0001-8777-2780} \and
Markus Stocker\inst{2}\orcidID{0000-0001-5492-3212} \and
S\"oren Auer\inst{2,1}\orcidID{0000-0002-0698-2864}}
\authorrunning{Jaradeh et al.}
\institute{L3S Research Center, Leibniz University of Hannover, Germany\\
    \email{jaradeh@l3s.de} \and
TIB Leibniz Information Centre for Science and Technology, Germany\\
    \email{\{markus.stocker,auer\}@tib.eu}}
\maketitle              % typeset the header of the contribution
\begin{abstract}
Answering questions on scholarly knowledge comprising text and other artifacts is a vital part of any research life cycle. 
Querying scholarly knowledge and retrieving suitable answers is currently hardly possible due to the following primary reason: machine inactionable, ambiguous and unstructured content in publications. 
We present JarvisQA, a BERT based system to answer questions on tabular views of scholarly knowledge graphs.
Such tables can be found in a variety of shapes in the scholarly literature (e.g., surveys, comparisons or results).
Our system can retrieve direct answers to a variety of different questions asked on tabular data in articles.
Furthermore, we present a preliminary dataset of related tables and a corresponding set of natural language questions.
This dataset is used as a benchmark for our system and can be reused by others. Additionally, JarvisQA is evaluated on two datasets against other baselines and shows an improvement of two to three folds in performance compared to related methods.

\end{abstract}

\section{Introduction}
\label{sec:intro}
Question Answering (QA) systems, such as Apple's Siri, Amazon's Alexa, or Google Now, answer questions by mining the answers from unstructured text corpora or open domain Knowledge Graphs (KG)~\cite{Karki2019QuestionModels}.
The direct applicability of these approaches to specialized domains such as scholarly knowledge is questionable.
On the one hand, no extensive knowledge graph for scholarly knowledge exists that can be employed in a question answering system.
On the other hand, scholarly knowledge is represented mainly as unstructured raw text in articles (in proceedings or journals)~\cite{Bornmann2015GrowthReferences}.
In unstructured artifacts, knowledge is not machine actionable, hardly processable, ambiguous~\cite{BosmanTheCommunication}, and particularly also not FAIR~\cite{Wilkinson2016TheStewardship}.
Still, amid unstructured information some semi-structured information exists, in particular in tabular representations (e.g., survey tables, literature overviews, and paper comparisons).
The task of QA on tabular data has challenges~\cite{Lin2002TheChallenges}, shared with other types of question answering systems.
We propose a method to perform QA specifically on scholarly knowledge graphs representing tabular data.
Moreover, we create a benchmark of tabular data retrieved from a scholarly knowledge graph and a set of related questions.
This benchmark is collected using the Open Research Knowledge Graph (ORKG)~\cite{Jaradeh2019OpenKnowledge}. 

The remainder of this article is structured as follows.
Section~\ref{ssec:example} motivates the work with an example.
Section~\ref{sec:rw} presents related work, which is supplemented by an analysis of the strengths and weaknesses of existing systems in the context of digital libraries.
Section~\ref{sec:approach} describes the proposed approach.
Section~\ref{sec:evaluation} presents the implementation and evaluation.
Section~\ref{sec:discussion-futurework} discusses results and future work.
Finally, Section~\ref{sec:conclusion} concludes the paper.

\subsubsection{Motivating Example}
\label{ssec:example}
\begin{figure}[tb]
    \centering
    \includegraphics[width=0.9\columnwidth]{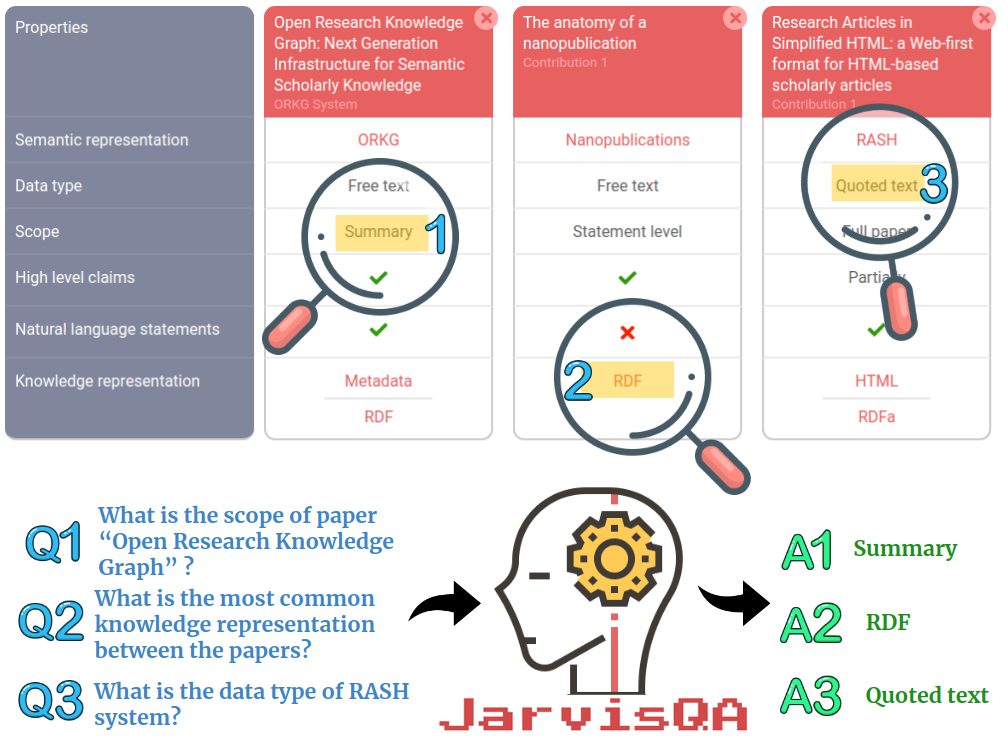}
    \caption{\textbf{Motivating Example.} \texttt{JarvisQA} takes as input a table of semi-structured information and tries to answer questions. Three types of questions are depicted here. (\textbf{Q1}) Answer is directly correlated with the question. (\textbf{Q2}) Aggregation of information from candidate results. (\textbf{Q3}) Answer relates to another cell in the table.}
    \label{fig:example}
\end{figure}
The research community has proposed many QA systems, but to the best of our knowledge none focus on scholarly knowledge.
Leveraging the ORKG~\cite{Jaradeh2019OpenKnowledge} and its structured scholarly knowledge, we propose a QA system specifically designed for this domain.
%In this case, the ORKG can be used as one source to tables describing scholarly information about specific papers.
Figure~\ref{fig:example} illustrates a tabular comparison view\footnote{\url{https://www.orkg.org/orkg/comparison/R8618}} of structured scholarly contribution descriptions.
Additionally, three questions related to the content of the comparison table are shown.
The answers are implicitly or explicitly provided in the cells of the table.
\texttt{JarvisQA} can answer different types of questions. For Q1, the answer has a direct correlation with the question. For Q2, the system should first find the ``knowledge representations'' in the table and then find the most common value.
For Q3, the answer is conditional upon finding another piece of information in the table first (i.e., JarvisQA has to find ``RASH'' in the table first), and then narrow its search to that column (or that paper) to find the correct answer.
%Table \ref{} presents the existing known table datasets that are used for training, evaluating, and developing question answering systems. To our knowledge no existing tables datasets is curated from the scholarly domain rather general purpose.

We tackle the following research questions: 

\begin{itemize}
    \item \textbf{RQ1:} \textit{Can a QA system retrieve answers from tabular representations of scholarly knowledge?}
    \item \textbf{RQ2:} \textit{What type of questions can be posed on tabular scholarly knowledge?}
    %\item \textbf{RQ3:} \textit{How can such a QA system be integrated in the context of digital libraries?}
\end{itemize}

\section{Related Work}
\label{sec:rw}
Question answering is an important research problem frequently tackled by research communities in different variations, applications, and directions.

In open domain question answering, various systems and techniques have been proposed that rely on different forms of background knowledge. 
Pipeline-based systems, such as OpenQA~\cite{Marx2014TowardsArchitecture}, present a modular framework using standardized components for creating QA systems on structured knowledge graphs (e.g., DBpedia~\cite{Auer2007DBpedia:Data}).
Frankenstein~\cite{Singh2018WhyTogether} creates the most suitable QA pipeline out of community created components based on the natural language input question.
QAnswer~\cite{Diefenbach2019QAnswer:End-users} is a multilingual QA system that queries different linked open data datasets to fetch correct answers.
Diefenbach et al.~\cite{Diefenbach2018CoreSurvey} discussed and compared other QA-over-KG systems (e.g., gAnswer~\cite{Zou2014NaturalApproach}, DEANNA~\cite{Yahya2012NaturalData}, and SINA~\cite{Shekarpour2015SINA:Data}) within the context of QALD ``Question Answering over Linked Data'' challanges~\cite{Lopez2013EvaluatingData}.

Other types of QA systems rely on the raw unstructured text to produce the answers.
Many of these systems are end-to-end systems that employ machine learning to mine the text and retrieve the answers.
Deep learning models (e.g., Transformers) are trained and fine-tuned on certain QA datasets to find the answers from within the text.
ALBERT~\cite{Lan2019ALBERT:Representations} is a descendent of BERT~\cite{Devlin2019BERT:Understanding} deep learning model.
At the time of writing, ALBERT holds the third top position in answering the questions of SQuAD \cite{Rajpurkar2016SQuad:Text}.
Such techniques model the linguistic knowledge from textual details and discard all the clutter in the text~\cite{Zinsser2012OnNonfiction}.
Other similar approaches include SG-Net \cite{Zhang2019SG-Net:Comprehension}, which uses syntax rules to guide the machine comprehension encoder-transformer models.

Tabular QA systems are also diverse and tackle the task with different techniques.
TF-IDF~\cite{RamosUsingQueries} is used to extract features from the tables and the question, and to match them.
Other models such as semantic parsers are used by Kwiatkowski et al.~\cite{KwiatkowskiScalingMatching} and Krishnamurthy and Kollar~\cite{Krishnamurthy2013JointlyWorld}.
Cheng et al.~\cite{Cheng2017LearningParsing} propose a neural semantic parser that uses predicate-argument structures to convert natural language text into intermediate structured representations, which are then mapped to different target domains (e.g., SQL).

Another category of table QA systems is neural systems.
TableQA~\cite{Vakulenko2017TableQA:Data} uses end-to-end memory networks to find a suitable cell in the table to choose. 
Wang et al.~\cite{WangATables} propose to use a directional self-attention network to find candidate tables and then use BiGRUs to score the answers.
Other table oriented QA systems include HILDB~\cite{Dua2013HindiSystem} that converts natural language into SQL.

In the plethora of systems that the community has developed over the past decade, no system addresses the scholarly information domain, specifically.
We propose a system to fill this gap and address the issues of QA on scholarly tabular data in the context of digital libraries (specifically with the ORKG\footnote{\url{https://orkg.org/}}).

%\subsubsection{Background Analysis}
%\label{ssec:analysis}
Though a variety of QA techniques exist, Digital Libraries (DL) primarily rely on standard information retrieval techniques~\cite{Schatz1997InformationNet}.
We briefly analyze and show when and how QA techniques can be used to improve information retrieval and search capabilities in the context of DLs.
Since DLs have different needs~\cite{Hersh2006InformationLibraries,Schatz1997InformationNet}; QA systems can improve information retrieval availability~\cite{Bloehdorn2007Ontology-basedLibraries}.
We argue that, Knowledge Graph based QA systems (or KG-QA) can work nicely within a DL context (i.e., aggregate information, list candidate answers).
Nevertheless, the majority of the existing scholarly KGs (such as MAG~\cite{Sinha2015AnApplications}, OC~\cite{Peroni2020OpenCitationsScholarship}) focus on metadata (e.g., authors, venues, and citations), not the scholarly knowledge content. 
%which is something a QA system can focus on but not different from a faceted search any digital library interface provides.
%Moreover, in the current state of the KGs addressing the content information, not the metadata, it would be rather useless to develop a QA system on that due to the lack of suitable data.

Another category of QA systems works on raw text, an important approach for DLs.
However, such systems are not fine-tuned on scholarly data; rather, they are designed for open domain data.
Furthermore, many of the end-to-end neural models have a built-in limitation~\cite{Yin2015NeuralAnswering} (i.e., model capacity) due to the architecture type, and as such cannot be used out of the box.
Some systems circumvent the problem of capacity (i.e., the inability to feed the model large amounts of text) by having a component of indexing (e.g., inverted index, concept and entity recognition) that can narrow down the amount of text that the system needs to process as the context for questions.

%Focusing on table QA systems, from the naming these systems use tables as their background knowledge.
%Tables are a common artefact in scholarly literature, so the adoption of this method can be more easily dealt with.
%Plus, these tables are still smaller in size than long raw text, so different kinds of QA techniques can be taken up to address the QA task.
%With our system, we focus on such kind of methods and evaluate the performance empirically to show the benefits of this system in the workflow of digital libraries compared to other standard information retrieval methods.

\section{Approach}
\label{sec:approach}
We propose a system, called \texttt{JarvisQA}, that answers Natural Language (NL) questions on tabular views of scholarly knowledge graphs, specifically tabular views comprising research contribution information from scientific articles.

\subsection{Data and Questions Collection}
\label{ssec:collection}
In order to evaluate our QA system we create the ORKG-QA benchmark, collected using the ORKG.
The ORKG provides structured comparisons~\cite{Oelen2020GenerateGraphs} of research contributions obtained from papers.
The ORKG-QA benchmark comprises a dataset that integrates 13 tables, covering information spanning more than 100 academic publications.
The data is collected through the ORKG API %\footnote{\url{https://www.orkg.org/orkg/api/classes/Comparison/resources/?items=999}},
and the featured set of tables\footnote{\url{https://www.orkg.org/orkg/featured-comparisons}}, which can be exported in CSV format.

Additionally, we created a set of questions that cover various types of information and facts that can be retrieved from those tables.
The benchmark consists of 80 questions in English.
The questions cover a variety of question types that can be asked in the context of tables in the scholarly literature.
These types of questions include aggregation questions (e.g., min, average and most common), ask questions (i.e., true, false), answer listing questions, and questions that rely on combining information.
In the ORKG-QA dataset\footnote{\url{https://doi.org/10.25835/0038751}}, 39\% are normal questions addressing individual cells in tables, 20\% are aggregation questions, 11\% are questions for which the answer relates to other parts of the table, and the rest are questions of different types (i.e., listings, ask queries, empty answers).

%% mention the other dataset
We also use the TabMCQ~\cite{Jauhar2016TabMCQ:Questions} QA dataset, specifically questions on the \textit{regents} tables.
TabMCQ was derived from multiple choice questions of 4th grade science exams and contains 39 tables and $3\,745$ related questions.
While TabMCQ is not a scholarly dataset, but is to the best of our knowledge the closest one available.
Since TabMCQ has only multiple-choice questions, we adapted the questions with only the correct choice. 
%Although this is not directly related to scholarly information, we include it to indicate the performance increase in our approach.

\subsection{JarvisQA system architecture}
\label{ssec:architecture}

\begin{figure}[tb]
    \centering
    \includegraphics[width=0.8\columnwidth]{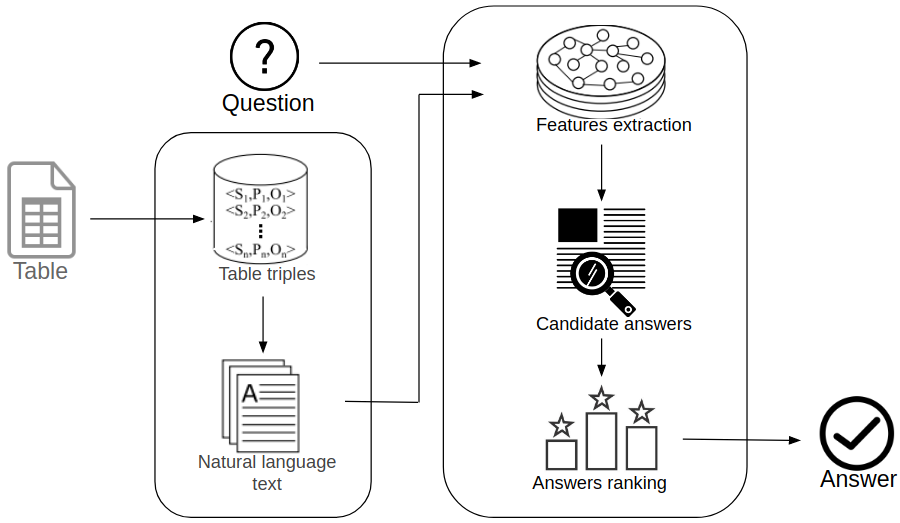}
    \caption{\textbf{System Architecture.} JarvisQA was designed with modularity in mind. The system has two main components. (a) \textbf{Table2Text (T2T)} component, which in turn has two functionalities: (1) to break the table into a set of triples $<s, p, o>$ and (2) to compile the triples into an NL sentence. Component (b) is the \textbf{engine of the QA system}, where an NL QA (BERT based) system is employed to answer the input question using the text, by extracting features, finding candidate answers, and ranking them.}
    \label{fig:architecture}
\end{figure}
\texttt{JarvisQA} is designed with modularity in mind.
Hence, the core QA components are replaceable with newer or more fine-tuned versions.
Figure~\ref{fig:architecture} depicts the architecture in more detail.
Since we used a natural language QA system, we need a pre-processing step that transforms the table information into the textual description (representing only the information contained in the table not the entire raw text of the article).
With the output of the ``Table2Text'' step and the input question, the NL QA system can reason over the question with the provided context (textual table description) and attempts to answer the question.
We now discuss the individual components of the architecture in more detail. 

\subsubsection{Table2Text (T2T) converter.}
Although \texttt{JarvisQA} operates on tabular data, the core QA engine processes textual contexts.
To that end, tables have to be converted into coherent text snippets that represent the entirety of the information presented in the table.
T2T component splits tables into its entries and converts entries into triples.
Table \ref{tab:table-example} illustrates a sample table containing some information about three publications, along with their triples and textual representations
%As an example, for the first entry (row) in the table the triples are presented. 
%\begin{lstlisting}[label={ls:triples}]
%<Paper1, hasSemanticRepresentation, ORKG>
%<Paper1, hasDataType, FreeText>
%<Paper1, hasScope, Summary>
%...
%\end{lstlisting}
compiled by the T2T component.
%\begin{lstlisting}[label={ls:t2t},basicstyle=\footnotesize  ]
%\textit{Paper 1's semantic representation is "ORKG", its data type is
%"Free Text", and its scope is "Summary"}.
%\end{lstlisting}
Furthermore, the T2T component enriches the textual description with aggregated information (i.e., max value of certain rows, most common value used within some columns). This enables the system to answer aggregation-type questions such as ``Which system has the maximum accuracy?'' and ``What is the most common method used among the papers?''.

\begin{savenotes}
\begin{table}[tb]
\scriptsize
\caption{\textbf{Sample of an input table.} The table is a part of the one shown in the motivating example.\protect\footnote{Fetched from \url{https://www.orkg.org/orkg/c/Zg4b1N}}
Below, the representation in triples and as text is displayed.}
\centering
%played a bit with the table layout ;) 
\begin{tabular}{%@{}cllll@{}}
    >{\centering\arraybackslash}m{2.2cm} %title
	>{\arraybackslash}m{2.7cm} % Semantic Representation
	>{\arraybackslash}m{2.5cm} % Data type
	>{\arraybackslash}m{2.5cm} % Scope 
	>{\arraybackslash}m{1.8cm} % High level claims
}
\toprule
\textbf{Title} & \textbf{Semantic \newline representation} & \textbf{Data type} & \textbf{Scope}  & \textbf{High level \newline claims} \\ \midrule
Paper 1~\cite{Jaradeh2019OpenKnowledge}        & ORKG                             & Free text          & Summary         & Yes                          \\
Paper 2~\cite{Groth2010TheNanopublication}        & Nanopublications                 & Free text          & Statement level & Yes                          \\
Paper 3~\cite{Peroni2017ResearchArticles}        & RASH                             & Quoted text        & Full paper      & Partially                  \\
%\bottomrule
\midrule
\end{tabular}
\begin{tabular}{%lm{10.8cm}}
    >{\centering\arraybackslash}m{4cm} %title
	>{\arraybackslash}m{8cm}
	}
\textbf{Triples} & 
$<$Paper1, hasSemanticRepresentation, ORKG$>$ \newline
$<$Paper1, hasDataType, FreeText$>$ \newline
$<$Paper1, hasScope, Summary$>$\newline
... \\
\textbf{Text} & Paper 1's semantic representation is "ORKG", its data type is
"Free Text", and its scope is "Summary" ... \\
\bottomrule
\end{tabular}
\label{tab:table-example}
\end{table}
\end{savenotes}

\subsubsection{QA core engine.}
This component is the primary building block of \texttt{JarvisQA}. It is where reasoning over questions happens.
The component uses a pre-trained natural language QA model.
The model is a deep transformer, fine tuned on the SQuADv2 dataset to perform the QA task.
The component is replaceable with any other similar transformer model (of different sizes and architectures).
Our base implementation uses a fine tuned version of a BERT model and we evaluate our model using different model sizes and architectures.
The model parameters are set: \textit{maximum sequence length} to 512, \textit{document stride} to 128, \textit{top k answers} to 10, \textit{maximum answer length} to 15, and the \textit{maximum question length} to 64.
As illustrated in Figure~\ref{fig:architecture}, the QA engine extracts sets of features from the questions and the text (i.e., embeddings), then it finds a set of candidate answers and ranks them by confidence score.
The benefits of such architecture are the flexibility in model choice, multilingualism, and reusability.
Different transformer models can replace ours to support other languages, other datasets, and potentially other features.
To accomplish these objectives, the system is built using the Transformers framework~\cite{Wolf2019HuggingFacesProcessing}.

\section{Experimental Study}
\label{sec:evaluation}

We empirically study the behavior of \texttt{JarvisQA} in the context of scholarly tables against different baselines.
The experimental setup consists of metrics and baselines.
Table~\ref{tab:metrics} lists the evaluation metrics for the performance measurements of the systems.
Since a QA system can produce multiple answers and the correct answer can be any of the retrieved answers we use a metric that takes the position of the answer into account.

\begin{table}[tb]
\caption{\textbf{Evaluation metrics} used to experimentally benchmark JarvisQA against other baselines.}
\scriptsize % you can play around with this if it is too small/big
\centering
\begin{tabular}{lp{10cm}}
\toprule
\textbf{Metric} & \textbf{Definition} \\ \midrule
\textit{Global Precision} & Ratio between correct answers retrieved in the top ranked position and the total number of questions. \\
\textit{Global Recall} & Ratio between the number of questions answered correctly at any position (here till the 10th retrieved answer) and the total number of questions. \\
\textit{F1-Score} & Harmonic mean of global precision and global recall. \\
Execution Time & Elapsed time between asking a question and returning the answer.\\
\textit{Inv. Time} & $1-\frac{average\ execution\ time\ for\ baseline}{maximum\ execution\ time\ for\ all\ systems}$. \\
\textit{In-Memory Size} & The total memory size used by system. \\
\textit{Inv. Memory} & $1-\frac{memory\ size\ of\ baseline}{maximum\ memory\ size\ among\ all\ systems}$ \\
\textit{Precision@K} & Cumulative precision at position K. \\
\textit{Recall@K} & Ratio of correctly answered questions in the top K position and total number of questions. \\
\textit{F1-Score@K} & Harmonic mean of precision and recall at position K. \\
\bottomrule
\end{tabular}
\label{tab:metrics}
\end{table}

\noindent As baselines we use the following two methods for answer generation: 

\begin{itemize}
    \item \textit{Random}: the answer is selected from all choices randomly.
    %It is used to measure whether the model gains the result on this task.
    \item \textit{Lucene}\footnote{\url{https://lucene.apache.org/}}: is a platform for indexing, retrieving unstructured information, and used as a search engine.
    We index the triple-generated sentences by Lucene. For each question, the top answer produced by Lucene is regarded as the final answer. 
    %We convert semi-structured tables into sentences, which can be indexed by Lucene. For each question, the top answer produced by Lucene is regarded as the final answer.
\end{itemize}

%\romannumeral 1) \textit{Random}: The answer is selected from all choices randomly. It is used to measure whether the model gains the result on this task.
%\romannumeral 2) \textit{Lucene}: Lucene\footnote{\url{https://lucene.apache.org/}} is a tool for indexing, retrieving unstructured information, and it is used as a search engine.
% We convert semi-structured tables into sentences, which can be indexed by Lucene. For each question, the top answer produced by Lucene is regarded as the final answer.

%\paragraph{\textbf{Implementation.}}
The evaluation was performed on an Ubuntu 18.04 machine with 128GB RAM and a 12 core Xeon processor.
The implementation is mostly based on HuggingFace Transformers\footnote{\url{https://github.com/huggingface/transformers}}, and is written in Python 3.7. 
The evaluation results for precision, recall, and F1-score are reproducible while
other metrics such as time and memory depend on the evaluation system hardware.
However, the ratio of the difference between the baselines should be similar or at least show a similar trend.
The code to reproduce the evaluation results and the presented results are available online.\footnote{\url{https://doi.org/10.5281/zenodo.3738666}}
%\footnote{\url{https://github.com/YaserJaradeh/JarvisQA}}.

\subsubsection{Experiment 1 - JarvisQA performance on the ORKG-QA benchmark.}
\label{ssec:exp1}
In order to evaluate the performance of \texttt{JarvisQA}, we run the system and other baselines on the ORKG-QA dataset at various $k$ values ($k$ denotes the position of the correct answer among all retrieved answers).
For this experiment we evaluate $k \in \left \{ 1,3,5,10 \right \}$.
Moreover, the experiment was conducted on a specific subset of questions (based on types) to show the performance of the system for certain categories of questions.
The tested question categories are: \textit{Normal}: normal questions about a specific cell in the table with a direct answer;
\textit{Aggregation}: questions about aggregation tasks on top of the table;
\textit{Related}: questions that require retrieving the answer from another cell in the table;
\textit{Similar}: questions that address the table using similar properties (e.g., synonyms).
Table~\ref{tab:exp1} shows the performance of the baselines and our system on the ORKG-QA benchmark.
The results show that \texttt{JarvisQA} performs better by 2-3 folds against Lucene, and Random baselines respectively.

\begin{table}[tb]
\caption{\textbf{JarvisQA performance on the ORKG-QA benchmark dataset} of tabular data. The evaluation metrics are precision, recall, and F1-score at $k$ position. JarvisQA is compared against two baselines on the overall dataset and specific question types. The symbol (-) indicates that the performance metric showed no difference than the reported value for higher $K$ values. The results suggest that JarvisQA outperforms the baselines by 2-3 folds.}
\centering
%\footnotesize
\begin{tabular}{@{}llcccclcccclcccc@{}}
\toprule
\multicolumn{1}{c}{\multirow{2}{*}{\textbf{\begin{tabular}[c]{@{}c@{}}Questions\\ type\end{tabular}}}} & \multicolumn{1}{c}{\multirow{2}{*}{\textbf{Baseline}}} & \multicolumn{4}{c}{\textbf{Precision @K}} &  & \multicolumn{4}{c}{\textbf{Recall @K}} &  & \multicolumn{4}{c}{\textbf{F1-Score @K}} \\ \cmidrule(lr){3-6} \cmidrule(lr){8-11} \cmidrule(l){13-16} 
\multicolumn{1}{c}{}                                                                                   & \multicolumn{1}{c}{}                                   & \#1      & \#3      & \#5      & \#10     &  & \#1      & \#3     & \#5     & \#10    &  & \#1      & \#3      & \#5      & \#10    \\ \midrule
All                                                                                                    & Random                                                 & 0.02     &  0.06        &  0.08        & 0.16         &  & 0.02     & 0.07        &  0.09       &  0.18       &  & 0.02     &  0.06        &  0.08        &  0.17       \\
All                                                                                                    & Lucene                                                 & 0.09     & 0.19     & 0.20      & 0.25     &  & 0.09     & 0.18    & 0.19    & 0.24    &  & 0.09     & 0.18     & 0.19     & 0.24    \\ \midrule
Normal                                                                                                 & JarvisQA                                                 & 0.41     & 0.47     & 0.55     & 0.61     &  & 0.41     & 0.47    & 0.53    & 0.61    &  & 0.41     & 0.47     & 0.54     & 0.61    \\
Aggregation                                                                                            & JarvisQA                                                 & 0.45     & -        & -        & -        &  & 0.45     & -       & -       & -       &  & 0.45     & -        & -        & -       \\
Related                                                                                                & JarvisQA                                                 & 0.50      & 0.50      & 1.00        & 1.00        &  & 0.50      & 0.50     & 1.00       & 1.00       &  & 0.50      & 0.500      & 1.00        & 1.00       \\
Similar                                                                                                & JarvisQA                                                 & 0.11     & 0.25     & 0.67     & -        &  & 0.11     & 0.25    & 0.67    & -       &  & 0.11     & 0.25     & 0.67     & -       \\ \midrule
All                                                                                                    & JarvisQA                                                 & 0.34     & 0.38     & 0.46     & \textbf{0.47}     &  & 0.35     & 0.38    & 0.46    & \textbf{0.48}    &  & 0.34     & 0.38     & 0.45     & \textbf{0.47}    \\ \bottomrule
\end{tabular}
\label{tab:exp1}
\end{table}

\subsubsection{Experiment 2 - Different models of QA and their performance.}
\label{ssec:exp2}
We evaluate different types of QA models simultaneously to show the difference in performance metrics, execution time, and resource usage.
Table~\ref{tab:exp2} illustrates the difference in performance on the ORKG-QA benchmark dataset for different classes of questions and the overall dataset.
\texttt{JarvisQA}'s QA engine employs the \texttt{BERT L/U/S2} model due to its execution time and overall higher accuracy at higher positions.
%\texttt{JarvisQA} makes use of \textit{BERT L/U/S2} as it's primary model.
%Though different models perform better on certain types of questions.
%We used \textit{BERT L/U/S2} as our base model due to its execution time and overall higher accuracy at higher positions. 

\begin{table}[tb]
\centering
\scriptsize
%\footnotesize
\caption{\textbf{Performance comparison of different deep learning models} on the task of question answering with different model sizes and architectures using the ORKG-QA benchmark dataset. The results suggest that different models perform differently on various question types, and generally the bigger the model the better it performs. For each question type, the best results are highlighted.}
\begin{tabular}{@{}clcccccccccccccc@{}}
\cmidrule(l){2-16}
\multirow{2}{*}{\textbf{}}                                                               & \multicolumn{1}{c}{\multirow{2}{*}{\textbf{\begin{tabular}[c]{@{}c@{}}Questions\\ type\end{tabular}}}} & \multicolumn{4}{c}{\textbf{Precision @K}} & \multicolumn{1}{l}{} & \multicolumn{4}{c}{\textbf{Recall @K}} & \multicolumn{1}{l}{} & \multicolumn{4}{c}{\textbf{F1-Score @K}} \\ \cmidrule(lr){3-6} \cmidrule(lr){8-11} \cmidrule(l){13-16} 
                                                                                         & \multicolumn{1}{c}{}                                                                                   & \#1       & \#3      & \#5      & \#10    &                      & \#1      & \#3     & \#5     & \#10    &                      & \#1      & \#3      & \#5      & \#10    \\ \midrule
\multirow{5}{*}{\textbf{\begin{tabular}[c]{@{}c@{}}BERT\\ L/U/S1\end{tabular}}}          & Normal                                                                                                 & 0.35      & 0.49     & 0.53     & \textbf{0.68}    &                      & 0.34     & 0.47    & 0.51    & \textbf{0.67}    &                      & 0.34     & 0.48     & 0.52     & \textbf{0.67}    \\
                                                                                         & Aggregation                                                                                            & 0.39      & 0.39     & 0.45     & -       &                      & 0.39     & 0.39    & 0.45    & -       &                      & 0.39     & 0.39     & 0.45     & -       \\
                                                                                         & Related                                                                                                & 0.50       & 0.64     & 0.64     & 0.80     &                      & 0.50      & 0.64    & 0.64    & 0.80     &                      & 0.50      & 0.64     & 0.64     & 0.80     \\
                                                                                         & Similar                                                                                                & 0.11      & 0.25     & \textbf{0.67}     & -       &                      & 0.11     & 0.25    & \textbf{0.67}    & -       &                      & 0.11     & 0.25     & \textbf{0.67}     & -       \\
                                                                                         & All                                                                                                    & 0.31      & 0.38     & 0.44     & \textbf{0.50}     &                      & 0.31     & 0.38    & 0.43    & \textbf{0.49}    &                      & 0.3      & 0.38     & 0.43     & \textbf{0.50}     \\ \midrule
\multirow{5}{*}{\textbf{\begin{tabular}[c]{@{}c@{}}BERT\\ L/C/S1\end{tabular}}}          & Normal                                                                                                 & 0.31      & 0.44     & 0.45     & -       &                      & 0.31     & 0.43    & 0.45    & -       &                      & 0.31     & 0.43     & 0.45     & -       \\
                                                                                         & Aggregation                                                                                            & 0.27      & 0.39     & 0.39     & 0.45    &                      & 0.29     & 0.39    & 0.39    & 0.45    &                      & 0.27     & 0.39     & 0.39     & 0.45    \\
                                                                                         & Related                                                                                                & 0.65      & \textbf{1.00}      & -        & -       &                      & 0.70      & \textbf{1.00}     & -       & -       &                      & 0.67     & \textbf{1.00}      & -        & -       \\
                                                                                         & Similar                                                                                                & 0.11      & 0.11     & 0.25     & 0.43    &                      & 0.11     & 0.11    & 0.25    & 0.43    &                      & 0.11     & 0.11     & 0.25     & 0.43    \\
                                                                                         & All                                                                                                    & 0.27      & 0.35     & 0.37     & 0.39    &                      & 0.29     & 0.37    & 0.39    & 0.41    &                      & 0.27     & 0.36     & 0.37     & 0.40     \\ \midrule
\multirow{5}{*}{\textbf{\begin{tabular}[c]{@{}c@{}}BERT\\ L/U/S2\end{tabular}}}          & Normal                                                                                                 & 0.41      & 0.47     & 0.55     & 0.61    &                      & 0.41     & 0.47    & 0.54    & 0.61    &                      & 0.41     & 0.47     & 0.54     & 0.61    \\
                                                                                         & Aggregation                                                                                            & 0.45      & -        & -        & -       &                      & 0.45     & -       & -       & -       &                      & 0.45     & -        & -        & -       \\
                                                                                         & Related                                                                                                & 0.50       & 0.50      & \textbf{1.00}      & -       &                      & 0.50      & 0.50     & \textbf{1.00}     & -       &                      & 0.50     & 0.50      & \textbf{1.00}      & -       \\
                                                                                         & Similar                                                                                                & 0.11      & 0.25     & \textbf{0.67}     & -       &                      & 0.11     & 0.25    & \textbf{0.67}    & -       &                      & 0.11     & 0.25     & \textbf{0.67}     & -       \\
                                                                                         & All                                                                                                    & 0.35      & 0.38     & 0.46     & 0.48    &                      & 0.35     & 0.38    & 0.46    & 0.48    &                      & 0.34     & 0.38     & 0.46     & 0.48    \\ \midrule
\multirow{5}{*}{\textbf{\begin{tabular}[c]{@{}c@{}}Distil\\ BERT\\ B/U/S1\end{tabular}}} & Normal                                                                                                 & 0.14      & 0.27     & 0.36     & 0.46    &                      & 0.16     & 0.29    & 0.36    & 0.46    &                      & 0.15     & 0.27     & 0.35     & 0.45    \\
                                                                                         & Aggregation                                                                                            & 0.22      & 0.39     & -        & -       &                      & 0.25     & 0.41    & -       & -       &                      & 0.24     & 0.39     & -        & -       \\
                                                                                         & Related                                                                                                & 0.31      & 0.50      & 0.64     & -       &                      & 0.31     & 0.50     & 0.64    & -       &                      & 0.31     & 0.50      & 0.64     & -       \\
                                                                                         & Similar                                                                                                & 0.00       & -        & -        & -       &                      & 0.00      & -       & -       & -       &                      & 0.00      & -        & -        & -       \\
                                                                                         & All                                                                                                    & 0.16      & 0.23     & 0.28     & 0.33    &                      & 0.17     & 0.26    & 0.29    & 0.35    &                      & 0.16     & 0.24     & 0.28     & 0.33    \\ \midrule
\multirow{5}{*}{\textbf{\begin{tabular}[c]{@{}c@{}}ALBERT\\ XL/S2\end{tabular}}}         & Normal                                                                                                 & 0.34      & 0.47     & 0.51     & -       &                      & 0.34     & 0.47    & 0.51    & -       &                      & 0.34     & 0.47     & 0.51     & -       \\
                                                                                         & Aggregation                                                                                            & 0.45      & 0.45     & \textbf{0.52}     & -       &                      & 0.45     & 0.45    & \textbf{0.52}    & -       &                      & 0.45     & 0.45     & \textbf{0.52}     & -       \\
                                                                                         & Related                                                                                                & \textbf{1.00}       & -        & -        & -       &                      & \textbf{1.00}      & -       & -       & -       &                      & \textbf{1.00}      & -        & -        & -       \\
                                                                                         & Similar                                                                                                & 0.43      & 0.43     & \textbf{0.67}     & -       &                      & 0.43     & 0.43    & \textbf{0.67}    & -       &                      & 0.43     & 0.43     & \textbf{0.67}     & -       \\
                                                                                         & All                                                                                                    & 0.36      & 0.42     & 0.46     & -       &                      & 0.37     & 0.43    & 0.47    & -       &                      & 0.36     & 0.42     & 0.46     & -       \\ \bottomrule
\end{tabular}
\label{tab:exp2}

B=Base; L=Large; XL=X-Large; C=Cased; U=Uncased; S1=Finetuned on SQuAD1;\newline S2=Finetuned on SQuAD2
\end{table}

\subsubsection{Experiment 3 - Trade-offs between different performance metrics.}
\label{ssec:exp3}
\begin{figure}[tb]
    \centering
    \includegraphics[width=0.75\textwidth]{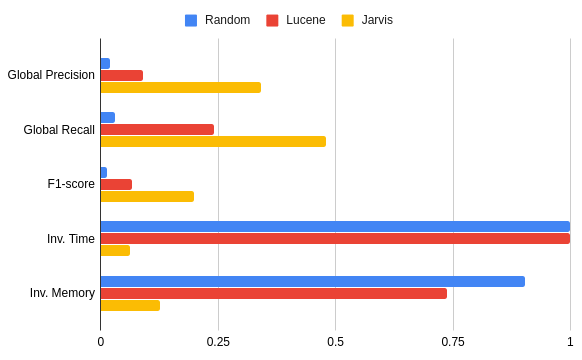}
    \caption{\textbf{Performance of the JarvisQA system}. JarvisQA and the baselines are compared in terms of Global Precision, Global Recall, Global F1-Score, Inv.Time, Inv.Memory; higher values are better. JarvisQA improves Precision, Recall, and F1-Score by up to three times at the cost of execution time and memory consumption.}
    \label{fig:trade-off}
\end{figure}

We illustrate trade-offs between different dimensions of performance metrics for the \texttt{JarvisQA} approach compared to the baselines.
We choose global precision, global recall, F1-score, in-memory size, and execution time as five different dimensions.
Figure~\ref{fig:trade-off} depicts the performance metrics trade-offs between our system and other baselines.
\texttt{JarvisQA} achieves higher precision and recall while consuming considerably more time and memory than the other baselines.

\subsubsection{Experiment 4 - Performance on TabMCQ.}
\label{ssec:exp4}
We also show the performance of our system on the TabMCQ dataset against the ORKG-QA dataset.
We see the same trend in both datasets, that \texttt{JarvisQA} outperforms the baselines by many folds.
TabMCQ is not directly related to scholarly knowledge. However, it shows that \texttt{JarvisQA} can generalize to related data and can answer questions about it.
Table~\ref{tab:exp4} presents the results of this experiment.

\begin{table}[tb]
\centering
\scriptsize %<---- table size here, replaced it with scriptsize
\caption{\textbf{Performance comparison using the two datasets TabMCQ and ORKG-QA} against JarvisQA and the baselines. The results suggest that JarvisQA outperforms the baselines by substantially on both datasets. Best results are highlighted for both datasets.}

\begin{tabular}{clcccccccccccccc}
\hline
\multirow{2}{*}{\textbf{System}} & \multicolumn{1}{c}{\multirow{2}{*}{\textbf{Dataset}}} & \multicolumn{4}{c}{\textbf{Precision @K}} & \multicolumn{1}{l}{} & \multicolumn{4}{c}{\textbf{Recall @K}} & \multicolumn{1}{l}{} & \multicolumn{4}{c}{\textbf{F1-Score @K}} \\ \cline{3-6} \cline{8-11} \cline{13-16} 
                                 & \multicolumn{1}{c}{}                                  & \#1      & \#3      & \#5      & \#10     &                      & \#1      & \#3     & \#5     & \#10    &                      & \#1      & \#3      & \#5      & \#10    \\ \hline
\multirow{2}{*}{\textbf{Random}} & TabMCQ                                                & 0.006    & 0.010     & 0.020     & 0.030     &                      & 0.010     & 0.020    & 0.030    & 0.040    &                      & 0.007    & 0.010     & 0.024    & 0.030    \\
                                 & ORKG                                                  & 0.020     & 0.060     & 0.080     & 0.160     &                      & 0.020     & 0.070    & 0.090    & 0.180    &                      & 0.020     & 0.060     & 0.080     & 0.017   \\ \hline
\multirow{2}{*}{\textbf{Lucene}} & TabMCQ                                                & 0.004    & 0.018    & 0.027    & 0.036    &                      & 0.006    & 0.017   & 0.026   & 0.037   &                      & 0.005    & 0.016    & 0.024    & 0.033   \\
                                 & ORKG                                                  & 0.090     & 0.190     & 0.200      & 0.250     &                      & 0.090     & 0.180    & 0.190    & 0.240    &                      & 0.090     & 0.180     & 0.190     & 0.240    \\ \hline
\multirow{2}{*}{\textbf{Jarvis}} & TabMCQ                                                & 0.060     & 0.090     & 0.100      & \textbf{0.110}     &                      & 0.070     & 0.090    & 0.110    & \textbf{0.120}    &                      & 0.060     & 0.080     & 0.100      & \textbf{0.110}    \\
                                 & ORKG                                                  & 0.340     & 0.380     & 0.460     & \textbf{0.470}     &                      & 0.350     & 0.380    & 0.460    & \textbf{0.480}    &                      & 0.340     & 0.380     & 0.450     & \textbf{0.470}    \\ \hline
\end{tabular}
\label{tab:exp4}
\end{table}

\section{Discussion and Future work}
\label{sec:discussion-futurework}
The main objective of \texttt{JarvisQA} is to serve as a system that allows users to ask natural language questions on tablar views of scholarly knowledge.
As such, the system addresses only a small part of the scholarly information corpus.

We performed several experimental evaluations to benchmark the performance of \texttt{JarvisQA} against other baselines using two different QA datasets.
Different datasets showed different results based on the types of questions and the nature of the scholarly data encoded in the tables.
Based on these extensive experiments, we conclude that usual information retrieval techniques used in search engines are failing to find specific answers for questions posed by a user.
\texttt{JarvisQA} outperforms the other baselines in terms of precision, recall, and F1-score measure at the cost of higher execution time and memory requirements.
Moreover, our system cannot yet answer all types of questions (e.g., non-answerable questions and listing questions).

Since \texttt{JarvisQA} utilizes a BERT based QA component, different components can perform differently, depending on the use case and scenario.
Our system struggles with answers spanning across multiple cells of the table, and also in answering true/false questions.
Furthermore, the answers are limited to information in the table (extractive method), since tables are not supplemented with further background information to improve the answers.

As indicated, the system can still be significantly improved. Future work will focus on improving answer selection techniques, and supporting more types of questions.
Additionally, we will improve and enlarge the ORKG-QA dataset to become a better benchmark with more tables (content) and questions.
\texttt{JarvisQA} currently selects the answer only from a single table, but use cases might require the combination of multiple tables or the identification of target table automatically (i.e., the system selects the table containing the correct answer from a pool of tables).
Moreover, in the context of digital libraries, we want to integrate the system into the ORKG infrastructure so it can be used on live data directly.

\section{Conclusion}
\label{sec:conclusion}
Retrieving answers from scientific literature is a complicated task. Manually answering questions on scholarly data is cumbersome, time consuming. %, and requires expertise.
Thus, an automatic method of answering questions posed on scientific content is needed.
\texttt{JarvisQA} is a question answering system addressing scholarly data that is encoded in tables or sub-graphs representing table content.
It can answer several types of questions on table content.
Furthermore, our ORKG-QA benchmark is a starting point to collaborate on adding more data to better train, evaluate, and test QA systems designed for tabular views of scholarly knowledge.
To conclude, \texttt{JarvisQA} addresses several open questions in current information retrieval in the scholarly communication domain, and contributes towards improved information retrieval on scholarly knowledge. 
It can help researchers, librarians, and ordinary users to inquire for answers with higher accuracy than traditional information retrieval methods.

\subsubsection*{Acknowledgments}
This work was co-funded by the European Research Council for the project ScienceGRAPH (Grant agreement ID: 819536) and the TIB Leibniz Information Centre for Science and Technology. The authors would like to thank our colleagues Kheir Eddine Farfar, Manuel Prinz, and especially Allard Oelen and Vitalis Wiens for their valuable input and comments.

\bibliographystyle{splncs04}
\bibliography{references}

\end{document}